\documentclass[preprint,12pt]{elsarticle}

\usepackage{amssymb}
\usepackage[utf8]{inputenc} 
\usepackage[T1]{fontenc}    
\usepackage{hyperref}
\usepackage{url}            
\usepackage{booktabs}       
\usepackage{amsfonts}       
\usepackage{nicefrac}       
\usepackage{microtype}      
\usepackage{lipsum}		
\usepackage{graphicx}
\usepackage{natbib}
\usepackage{doi}
\usepackage{subfigure}
\usepackage[linesnumbered, ruled]{algorithm2e}
\usepackage{algpseudocode}
\usepackage{amsmath}
\usepackage{bm}
\usepackage{float}
\usepackage{wrapfig} 

\journal{elsevier}

\begin{document}

\begin{frontmatter}



\title{Self-attention-based multi-agent continuous control method in cooperative environments}


\author[inst1]{Kai Liu}

\affiliation[inst1]{organization={School of Mechanical and Electrical Engineering},
            addressline={University of Electronic Science and Technology of China}, 
            city={Chengdu},
            postcode={611731}, 
            country={China}}

\author[inst1]{Yuyang Zhao}
\author[inst2]{Gang Wang}
\author[inst1]{Bei Peng\corref{mycorrespondingauthor}}
\cortext[mycorrespondingauthor]{Corresponding author}
\ead{beipeng@uestc.edu.cn}

\affiliation[inst2]{organization={School of Information and Communication Engineering},
            addressline={University of Electronic Science and Technology of China}, 
            city={Chengdu},
            postcode={611731}, 
            country={China}}

\begin{abstract}
Cooperative problems under continuous control have always been the focus of multi-agent reinforcement learning. Existing algorithms suffer from the problem of uneven learning degree with the increase of the number of agents.
In this paper, a new structure for a multi-agent actor critic is proposed, and the self-attention mechanism is applied in the critic network and the value decomposition method used to solve the uneven problem.
The proposed algorithm makes full use of the samples in the replay memory buffer to learn the behavior of a class of agents.
First, a new update method is proposed for policy networks that promotes learning efficiency.
Second, the utilization of samples is improved, at the same time reflecting the ability of perspective-taking among groups.
Finally, the "deceptive signal" in training is eliminated and the learning degree among agents is more uniform than in the existing methods. 
Multiple experiments were conducted in two typical scenarios of a multi-agent particle environment. Experimental results show that the proposed algorithm can perform better than the state-of-the-art ones, and that it exhibits higher learning efficiency with an increasing number of agents.
\end{abstract}




\end{frontmatter}


\section{Introduction}
\label{sec:1}
Multi-agent deep reinforcement learning (MARL) has made excellent progress in recent years, and it is considered to be an important component in constructing general artificial intelligence (AI)\citep{survey-critique-marl,21reward}.
MARL can be divided into discrete problems and continuous problems according to the action space.
With the proposal of the StarCraft II learning environment (PySC2) and the StarCraft Multi Agent Challenge (SMAC)\citep{pysc2-nature,smac}, algorithms that can be used for environments with discrete action spaces have been fully developed in recent years, especially for solving large-scale problems\citep{vdn,q-mix,qtran}.
However, in the field of continuous action spaces, which is widely used in robot navigation\citep{uav-coninuous,robot}, cluster control\citep{maddpg-flocking,maddpg-swarm}, power systems\citep{power,power-ddpg}, and games\citep{mappo,parameter-sharing}, few studies focus on large-scale continuous control problems.
In this paper, the focus is on MARL in a cooperative environment with continuous actions spaces, and an attempt is made to address the problems of a large number of agents.

What accompanies multi-agent tasks are a decentralized partially observable Markov decision process (DEC-POMDP)\citep{dec-pomdp} and a non-stationary environment. Traditional reinforcement learning approaches such as  Q-learning\citep{dqn-nature,DQN} and policy gradients\citep{vpg} are poorly suited to multi-agent environments. The generally accepted framework for MARL is based on the actor-critic structure, in which the policy network can independently obtain local information and output actions based on its own partial observations, and the action-value function integrates the observations and actions of all of the agents to evaluate the current agent's action. The set of observations and actions of all of the agents at a certain moment is regarded as a state of the POMDP, which solves the non-stationary problem of the environment. This structure, combined with centralized training and decentralized execution (CTDE), such as MADDPG\citep{maddpg}, is the most widely used structure in the field of multi-agent continuous control.

However, with the deepening of research, the following two disadvantages of this framework have been revealed.\\
(1) The input of the action-value function has a significant amount of redundant information and the action space grows exponentially with an increasing number of agents.\\
(2) The existing methods perform worse with the increasing number of agents, and most rewards offered by the environment are not related to the individual agent, which causes the problem of uneven learning degree among agents.\\

The first problem has been mentioned by many researchers, and many new modeling methods have emerged\citep{ATT-MADDPG,DAACMP,MAAC,MACAAC,redundant-communicate,redundant-individualized,redundant-schedule-comm}, but the second problem has rarely been studied.
It is considered herein that there is a "deceptive signal" generated by the second problem and a new actor-critic structure is thus offered to solve this problem.

The two problems mentioned above were reconsidered, and it was noted that the core of the multi-agent problem is to deal with the role of the individual agent in a multi-agent environment. It is best for the action of one agent to be determined by considering the states of other agents comprehensively.
This is similar to the natural language processing (NLP) problem, in which a word must be contextualized to determine its own meaning.
The similarity between MARL and NLP problems provided the motivation to transplant the self-attention mechanism\citep{transformer} into the multi-agent field, which is the state-of-the-art (SOTA) to deal with the semantic problems of words in different sentences.

To combine the self-attention mechanism with an actor-critic structure, three techniques are proposed that are  organized around the self-attention mechanism to maximize its effectiveness.\\
(1) The paralleled connection of observation and action of all of the agents is used as the new input structure.\\
(2) The value-decomposition approach\citep{vdn} is applied to learn an optimal linear value decomposition from the joint reward. This avoids the spurious reward signals that are generated from independent learners.\\
(3) A one-to-all policy update mode is introduced to accelerate training. With the help of double-Q learning, target-policy smoothing, and "delayed" policy updates\citep{td3,matd3}, this technique can improve the training speed while maintaining stability.\\
Most importantly, the self-attention-based action-value function deals with the problem of information fusion in an efficient way and achieves perspective-taking among agents.

In this paper, a new approach called SA-MATD3 is proposed that combines the self-attention mechanism with the MATD3 algorithm in an efficient way. The double-attention structure-based algorithm (DSA-MATD3) is also derived, which cannot perform decentralized execution, but it generally works better and faster than decentralized execution algorithms; thus, use of DSA-MATD3 is recommended in projects that do not strictly require distributed execution.

The rest of this paper is organized as follows: In Section 2, the relevant background is introduced. The SA-MATD3 algorithms are derived in Section 3, and in Section 4, the experimental settings and results are introduced. Section 5 concludes the paper.\\

\section{Background}
\label{sec:3}
\subsection{Key variables}
All of the commonly used variables in this paper are listed in Table \ref{tab:T1} for the reader's convenience. For ease of recall, subscripts represent the time step and superscripts represent the agent number; an overtilde represents the next state or target policy.\\

\begin{table}[H]
	\caption{Key variables}
	\centering
	\resizebox{\textwidth}{!}{
	\begin{tabular}{ll}
		\toprule
		Symbols     & Description \\
		\midrule
	    ${\bf{S}} = [{s_1},{s_2},...,{s_t}]$ & Set of current states at all times \\
		${\bf{\tilde S}} = [{\tilde s_1},{\tilde s_2},...,{\tilde s_t}],{\tilde s_t} = {s_{t + 1}}$ & Set of next states at all times \\
		${\bf{A}} = [{A^1},{A^2},...,{A^n}]$ & Set of actions of each agent \\
		${A^i} = [a_1^i,a_2^i,...,a_t^i]$ & Set of actions of $i$th agent at all moments \\
		${{\bf{a}}_j} = [a_j^1,a_j^2,...,a_j^n]$ & Set of actions of all agents at time $j$\\
		${\bf{T}}({{\tilde s}_j}|{s_j},{{\bf{a}}_j}):{\bf{S}} \times {\bf{A}} \times {\bf{\tilde S}}$ & State-transition equation\\
		${\bf{O}} = [{O^1},{O^2},...,{O^n}]$ & Joint observation of all agents\\
		${\bf{\tilde O}} = [{{\tilde O}^1},{{\tilde O}^2},...,{{\tilde O}^n}]$ & Joint next observation of all agents\\
		${O^i} = [o_1^i,o_2^i,...,o_t^i]$ & Set of observations of $i$th agent at all moments\\
		$r = [{r_1},{r_2},...,{r_t}]$ & Set of joint rewards\\
		$\gamma  \in [0,1]$ & Discount factor in Bellman equation\\
		${{\bf{\pi }}_{{\theta ^i}}},{{\bf{\pi }}^i}$ & Policy network of $i$th agent with parameter ${\theta ^i}$\\
		${{\bf{\pi }}_{{{\tilde \theta }^i}}},{{{\bf{\tilde \pi }}}^i}$ & Target policy network of $i$th agent with delayed parameter ${\tilde \theta ^i}$\\
		${Q_{{{\bf{\pi }}^i}}},{Q_{{{{\bf{\tilde \pi }}}^i}}}$ & Action-value function and target action-value function of $i$th agent\\
		${Q_{\bf{\pi }}},{Q_{{\bf{\tilde \pi }}}}$ & Centralized action-value function and centralized target action-value function\\
		${{\tilde A}^i}$ & Target action of $i$th agent.\\
		${{\bar A}^i}$ & Target action with smoothing noise of $i$th agent\\
		\bottomrule
	\end{tabular}
	}
	\label{tab:T1}
\end{table}

\subsection{Dec-POMDP}
Multi-agent setting in this paper can be formulated as Dec-POMDPs, which are formally defined as a tuple $({\bf{S}},{\bf{A}},{\bf{T}},r,{\bf{O}},Z,\gamma )$, where ${\bf{S}}$ is the set of states $s_j$, ${\bf{A}}$ the set of actions ${A^i}$ of each agent, ${A^i}$ the set of individual actions $a_j^i$ at all moments, ${\bf{T}}({\tilde s_j}|{s_j},{{\bf{a}}_j}):{\bf{S}} \times {\bf{A}} \times {\bf{\tilde S}}$ the state transition equation, ${\bf{O}} = [{O^1},{O^2},...,{O^n}]$ the set of joint observations ${O^i} = [o_1^i,o_2^i,...,o_t^i]$, which is determined by the observation equation $Z:{\bf{S}} \times {\bf{A}} \to {\bf{O}}$, and $\gamma  \in [0,1]$ is the discount factor in the Bellman equation.

At time $j$, each agent takes an action $a_j^i$ based on its current observation $o_j^i$, resulting a new state $s{'_j}({s_{j + 1}})$, and the environment gives a reward ${r_j}$ according to actions of all agents at this time. Each agent tries to learn a policy ${{\bf{\pi }}^i}({O^i}|{A^i})$ to maximize the expectancy reward ${\rm E}(R)$, where $R = \sum\nolimits_1^T {{\gamma _j}} {r_j}$ represents the weighted sum of reward ${r_j}$ at each moment and $T$ the time limit of the entire learning process.

In this paper, it is assumed that the environment can be fully observed jointly: ${\bf{S}} \buildrel \Delta \over = {\bf{O}},{\bf{\tilde S}} \buildrel \Delta \over = {\bf{\tilde O}}$, and the reward ${r_j}$ is a joint reward, which is given equally to the same type\footnote{The same type means a group of agents that perform the same job, although their functions may be different.} of agents; the above two points are a consensus of DEC-POMDPs.
\subsection{MADDPG}
The MADDPG algorithm is the most widely used MARL algorithm in the continuous control field. The main idea is to use the CTDE method combined with the actor-critic architecture to deal with the variety of multi-agent tasks. The critic is designed to use information of all of the agents, and thus can be used for a mixed cooperative-competitive environment. The MADDPG algorithm is an off-policy method, which applies the experience replay and target network to stabilize training, and it has high data efficiency in multi-agent problems compared with on-policy algorithms.

Considering an $N$ continuous policy ${{\bf{\pi }}_{{\theta ^i}}}$(abbreviated as ${{\bf{\pi }}_i}$) with parameters ${\theta ^i}$, and that the experience replay buffer $D$ contains a  tuple $({\bf{S}},{\bf{\tilde S}},{\bf{A}},r)$ that records all of the agent's experiences, the gradient can be written as
\begin{equation}
	{\nabla _{{\theta ^i}}}J({{\bf{\pi }}_i}) = {{\mathop{\rm E}\nolimits} _{{\bf{S}},{\bf{A}} \sim D}}[{\nabla _{{\theta ^i}}}{{\bf{\pi }}^i}({A^i}|{O^i}){\nabla _{{A^i}}}{Q_{{{\bf{\pi }}^i}}}({\bf{S}},{A^1},{A^2},...,{A^n}){|_{{A^i} = {{\bf{\pi }}^i}({O^i})}}].
\end{equation}
The action-value function can be updated according to the following equation:
\begin{equation}
    \begin{array}{l}
    L({\theta _i}) = {{\mathop{\rm E}\nolimits} _{{\bf{S}},{\bf{\tilde S}},{\bf{A}},{\bf{R}}\relax D}}[{({Q_{{{\bf{\pi }}^i}}}({\bf{S}},{\bf{A}}) - y)^2}],\\
    y = r + \gamma (1 - d){Q_{{{{\bf{\tilde \pi }}}^i}}}({\bf{\tilde S}},{{\tilde A}^1}, \cdots ,{{\tilde A}^n}){|_{{{\tilde A}^i} = {{{\bf{\tilde \pi }}}^i}({{\tilde O}^i})}},
    \end{array}
\end{equation}
where $\{ {{\bf{\pi }}_{{{\tilde \theta }^1}}}, \cdots ,{{\bf{\pi }}_{{{\tilde \theta }^n}}}\}$, abbreviated as $\{ {{\bf{\tilde \pi }}_1}, \cdots ,{{\bf{\tilde \pi }}_n}\} $, is a set of target policies with delayed parameters ${\tilde \theta ^i}$.
\subsection{MATD3}
\label{sec:matd3}
While the MADDPG algorithm can sometimes achieve great performance, it is frequently fragile with respect to hyper-parameters and other kinds of tunings. A common failure mode for the MADDPG algorithm is that the learned Q-function begins to dramatically overestimate Q values, which then leads to the policy breaking, because it exploits the errors in the Q-function. The multi-agent twin delayed DDPG (MATD3) is an algorithm that addresses this issue by introducing three critical tricks.\\
1. Clipped Double-Q Learning. MATD3 learns two Q functions and uses the smaller of the two Q values to form the target Bellman equation, which relieves the problem of overestimation:
\begin{equation}
    y = r + \gamma (1 - d)\min {Q_{{\bf{\tilde \pi }}_l^i}}(\tilde O,{{\bar A}^1}, \cdots ,{{\bar A}^n}),l = 1,2.
\end{equation}
2. Target-Policy Smoothing. The update of the critic is smoothed by adding a tiny noise to target actions, to make it harder for the policy to exploit Q-function errors. Here, $N(0,e)$ represents the Gaussian noise with $0$ as the mean and $e$ as the variance; ${{\bf{a}}_L},{{\bf{a}}_H}$ is the lower and upper bound, respectively, of the smoothed action:
\begin{equation}
    \bar A = clip({{\bf{\tilde \pi }}^i}({\tilde O^i}) + \widetilde N(0,e),{{\bf{a}}_L},{{\bf{a}}_H}).
\end{equation}
3. “Delayed” Policy Updates. MATD3 updates the policy less frequently than the Q-function, usually every two updates of the Q-function accompanied by one update of the policy, which makes the critic networks converge in advance and updates the actor network with a more reasonable gradient.
\subsection{Self-attention mechanism}
The self-attention mechanism was originally proposed in Transformer\citep{transformer}, which has made breakthrough achievements in the fields of NLP and computer vision (CV) in recent years. It can effectively deal with the semantics of words in sentences, compared with long-short term memory (LSTM) and recurrent neural networks(RNNs), and it supports parallel computing and has higher efficiency. In NLP, the input matrix is $X \in {R^{Batch \times T \times {D_{in}}}}$ containing a $Batch$ sentence, and each sentence contains $T$ words and each word has ${D_{in}}$ as the maximum length. The attention process can be expressed as the matrix multiplication of query, key, and value:
\begin{equation}
    \begin{array}{l}
    att(X) = softmax(A)X{W_{val}},\\
    A = X{W_{qry}}W_{key}^{\rm T}{X^{\rm T}}.
    \end{array}
\end{equation}

\begin{figure}[H]
    \centering
    \includegraphics[width=0.7\textwidth]{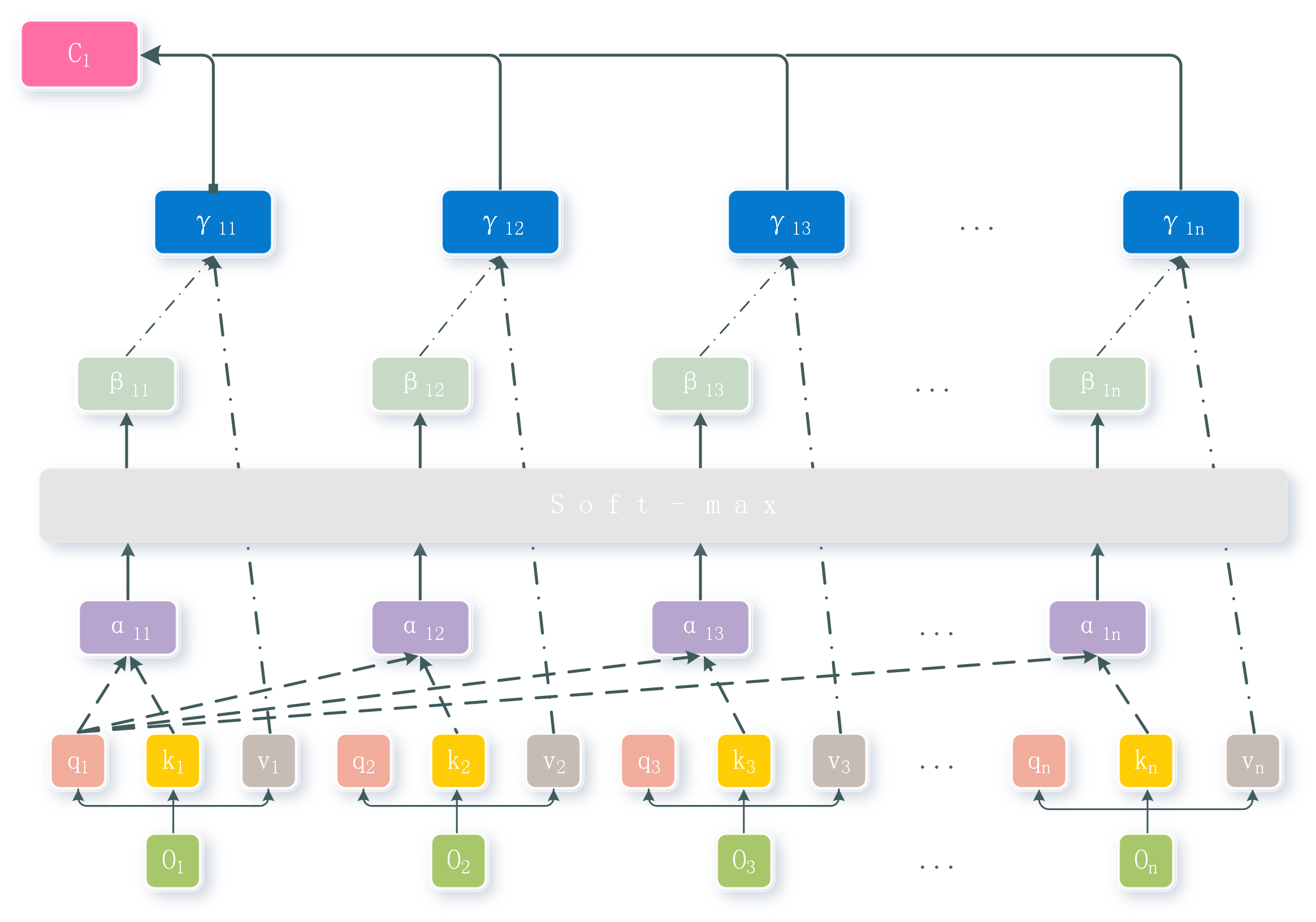}
    \caption{Schematic of single-head self-attention structure, where solid lines represent linear neural network calculation and dotted lines represent matrix multiplication. ${O_i}$ is input of size ${D_{in}}$, ${a_{i,j}}$ the $i$th row and $j$th column of matrix $A$, ${\gamma _{i,j}}$ the $i$th row and $j$th column of matrix $att(X)$, and, finally, $C_i$ is the sum of rows to obtain the output value of each input through the attention structure.}
    \label{fig:attention}
\end{figure}

The linear layers that are used to generate [query,key,value] are represented as ${W_{qry}} \in {R^{{D_{in}} \times {D_k}}}, {W_{key}} \in {R^{{D_{in}} \times {D_k}}}, and {W_{val}} \in {R^{{D_{in}} \times {D_{out}}}}$, respectively, and the diagram of the self-attention structure is shown in Figure \ref{fig:attention}. To extract more correlation features, researchers usually use a multi-head attention mechanism to generate multiple $W_{qry}^i,W_{key}^i,W_{val}^i$ in parallel, and finally maintain dimensional consistency through a linear layer ${W_{out}}$:
\begin{equation}
    X = \mathop {concat}\limits_{i \in [{N_h}]} [at{t^i}(X)]{W_{out}}.
\end{equation}

The consistency of input and output dimensions through the self-attention layer enables them to be stacked one by one. Usually, the encoder or decoder is composed of a stack of $N$ identical self-attention layers. To facilitate training, a residual\citep{residual} is used to prevent the gradient from disappearing and normalization method\citep{layer-norm} is used to accelerate training.

Figure \ref{fig:attention} illustrates the perspective-taking proposed in this paper from a microscopic perspective. If one changes the order of ${O_1}$ and ${O_2}$, the output ${C_1}$ and ${C_2}$, respectively, will still correspond to ${O_1}$ and ${O_2}$, respectively, and the result will remain the same, which means that the input of all of the agents will be trained and treated equally in self-attention layers.

\section{Method}
\label{sec:4}
\subsection{New input structure for critic network}
\begin{figure}[H]
    \centering
    \includegraphics[width=0.8\textwidth]{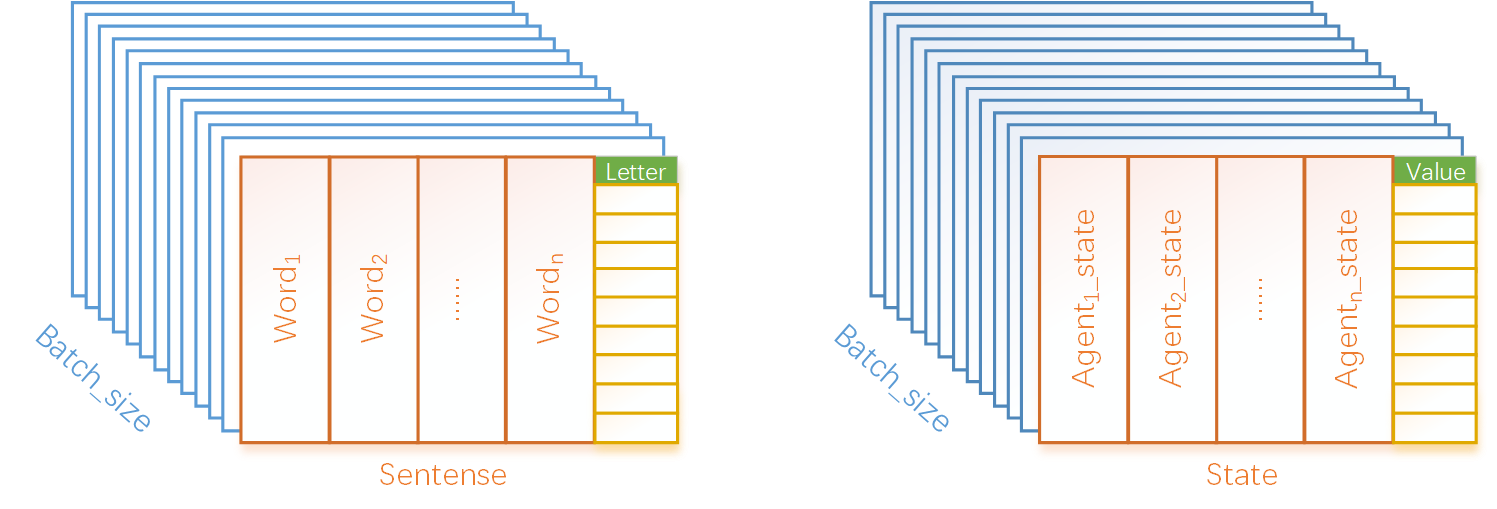}
    \caption{New input structure for critic network that is similar to that for NLP training. Left, NLP input structure; right, new input structure for MARL.}
    \label{fig:dimension}
\end{figure}

The self-attention mechanism cannot be directly applied to multi-agent reinforcement learning because they have different input dimensions. The traditional algorithms use observation and action concatenation of all of the agents as the input of critic network, which is a relatively rough way to incorporate other agents. To meet the dimensional demand of a self-attention network, observation and action of each agent are connected in parallel at first. 
Compared with existing MARL algorithms, in which the input vector is represented as $[batch-size,all-states]$, the dimension of the new input structure should be expressed as $[batch-size,agent-num,agent-state]$. The modified input structure is shown in Figure \ref{fig:dimension}, which is similar to the input for NLP training. This structure shows the inherent relevance of the multi-agent and NLP problems. In NLP, a word must be contextualized to determine its own meaning. A similar idea is also reflected in multi-agent tasks: one must deal with the role of a single agent in a multi-agent environment, and it is better for the action of this agent to be determined by considering the states of other agents comprehensively. Attention was devised to deal with the semantic problems of words in different sentences, and it also works for multi-agent tasks.

\subsection{Total Q for critic update}
The problem of uneven learning degree in a cooperative environment was studied and it was found that the Q value calculated by the current algorithm is not equal to the joint reward used for updating. In the Bellman equation, the Q value one wants to update is generated by the critic network of each agent, and the reward used to update this Q value is the joint reward provided by the environment. Taking the predator-prey experimental environment as an example, it is assume that there are three predators and one prey. According to the consensus of DEC-POMDP, when any predator collides with the prey, all of the predators will get the joint reward. Now, the situation when predator 2 collides with the prey, and predator 1 does not do anything worthwhile at this moment, is considered. When the critic network of predator 1 uses this sample for training, its Q value is updated by the joint reward, which gives it a "deceptive signal" that this positive reward may be brought about by its action. When the number of agents is large, this kind of "deceptive signal" will increase exponentially with increasing number of agents. Therefore, the performance of the existing multi-agent continuous control algorithms will perform poorly with increasing number of agents. The reason is that each agent does not have a Q value for its own behavior, and the Q values calculated by the centralized critic network are vague and ambiguous. In this paper, the new input structure and self-attention mechanism is adopted for the critic network. Through forward propagation, the corresponding Q value of each agent can be obtained simultaneously, and the idea of value decomposition is adopted and the Q values of each same type of agent are added to obtain a total Q value. This total Q value has a stronger correspondence with the joint reward and is more reasonable for updating.

\subsection{SA-MATD3}
\begin{figure}[H]
    \centering
    \includegraphics[width=0.8\textwidth]{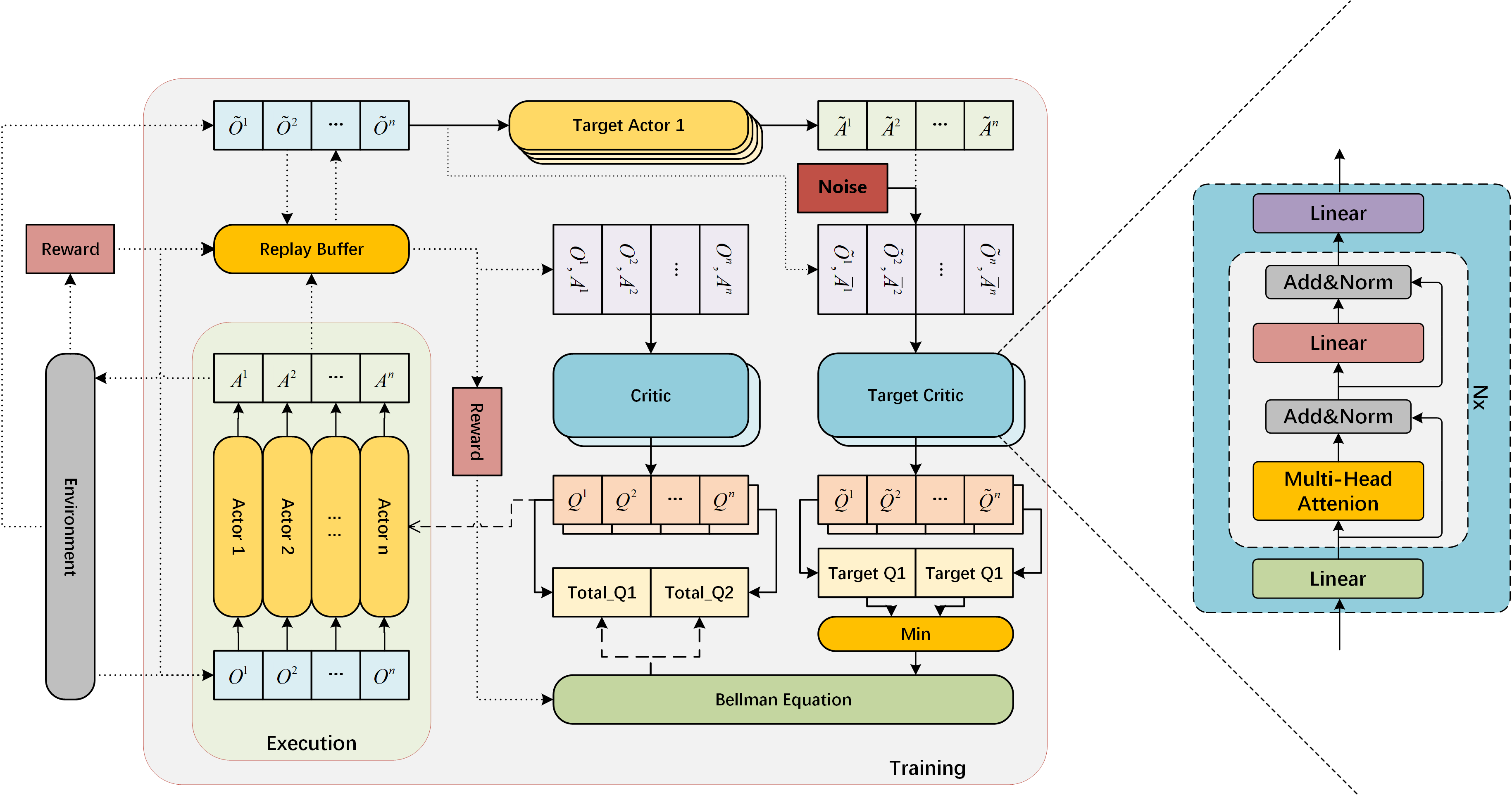}
    \caption{Diagram of SA-MADDPG. Dotted line represents the transmission of data, solid line the computation of the neural network, and dashed line the gradient update.}
    \label{fig:sa-matd3}
\end{figure}

In SA-MATD3, each agent has its own actor network that outputs the action of each agent using the individual observation; the same type of agent shares one critic network, and the actor network is the most commonly used multilayer perceptron (MLP). The critic network uses a module of the self-attention mechanism that takes the new structure as input and outputs Q values of each agent at the same time. The positional encoding among the same type of agents was removed, and thus the same type of agent could be treated and trained in the same way. The existence of positional encoding breaks the perspective-taking ability among a group of agents. The idea of value decomposition is adopted and the Q values of the same type of agents are added as the total Q value, which is the key to solving the problem of uneven learning degree among agents. SA-MATD3 also adopts the three critical tricks to address the overestimation problem that were mentioned in Section \ref{sec:matd3}, and will not be repeated here. The overall approach of SA-MATD3 is clearly illustrated in Figure \ref{fig:sa-matd3}.

Taking ${{\bf{\pi }}_{{\theta ^i}}}$ as the policy of the current agent with ${\theta _i}$ as the parameter, abbreviated as ${{\bf{\pi }}^i}$, a series of random samples $[{O^1},{O^2},...,{O^n},{A^1},{A^2},...,{A^n},r,{\tilde O^1},{\tilde O^2},...,{\tilde O^n}]$ are taken from the replay buffer $D$.
The update of the action-value function can be represented as
\begin{equation}
    \begin{array}{l}
y = r + \min {Q_{{{{\bf{\tilde \pi }}}^{1,2}}}}(({{\tilde O}^1},{{\bar A}^1}),({{\tilde O}^2},{{\bar A}^2}),...,({{\tilde O}^n},{{\bar A}^n})){|_{{{\bar A}^i} = {{{\bf{\tilde \pi }}}^i}({{\tilde O}^i}) + \widetilde N(0,{e_{critic}})}},\\
{\phi ^{1,2}} \leftarrow argmi{n_{{\phi ^{1,2}}}}\sum {{{({Q_{{{\bf{\pi }}^{1,2}}}}(({O^1},{A^1}),({O^2},{A^2}),...,({O^n},{A^n})) - y)}^2}}. 
\end{array}
\end{equation}

The "delayed" policy updates are adopted and the update of the policy network can be expressed as
\begin{equation}
    {\nabla _{{\theta ^i}}}J({{\bf{\pi }}_i}) = {{\mathop{\rm E}\nolimits} _{{\bf{O}},{\bf{A}} \sim D}}[{\nabla _{{\theta ^i}}}{{\bf{\pi }}^i}({A^i}|{O^i}){\nabla _{{A^i}}}{Q_{\bf{\pi }}}(({O^1},{A^1}),({O^2},{A^2}),...,({O^n},{A^n})){|_{{A^i} = {{\bf{\pi }}^i}({O^i})}}].
\end{equation}

\begin{figure}[H]
    \centering
    \includegraphics[width=0.8\textwidth]{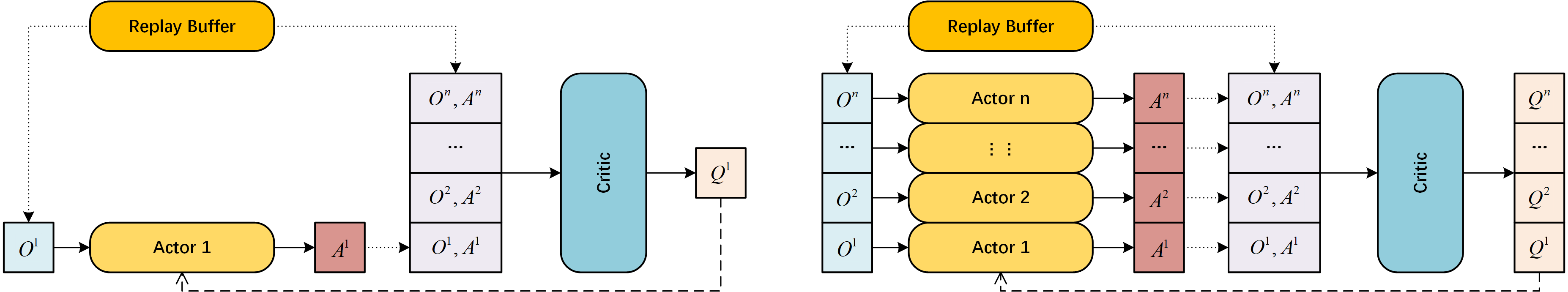}
    \caption{Different ways of executing policy updates. Left, one-to-one mode in MADDPG and MATD3; right, one-to-all mode in SA-MATD3.}
    \label{fig:policy_update}
\end{figure}

In the proposed algorithm, the policy of all of the agents are updated at the same time, which is called in this work the one-to-all mode for policy update. Since one critic network is shared among the same type of agents, when calculating the value function the actions calculated by all policy networks are used, while in the MADDPG algorithm most of the actions are taken from the replay memory buffer. This modification allows all the policy networks to obtain the gradient after one calculation of the action-value function, so the policies of all of the agents can be updated simultaneously. With the use of the self-attention mechanism, double-Q learning, and "delayed" policy updates, this radical approach could still maintain stability while obtaining high learning efficiency at the same time. The different methods of policy updating between the proposed and existing algorithms are shown in Figure \ref{fig:policy_update}.

Here, SA-MATD3 is rendered completely with pseudo-code.\\
\begin{algorithm}[H]
\caption{Self-attention-based multi-agent twin delayed deep deterministic policy gradient.}
  Initialize actor network ${{\bf{\pi }}^i}$ for each agent $i$ with random parameters ${\theta ^i}$, initialize two critic network ${Q_{{{\bf{\pi }}^1}}},{Q_{{{\bf{\pi }}^2}}}$ with parameters ${\phi ^1},{\phi ^2}$, initialize target actor networks for each agent $i$, ${\tilde \theta ^i} \leftarrow {\theta ^i}$, initialize target critic networks ${\tilde \phi ^{1,2}} \leftarrow {\phi ^{1,2}}$,initial replay buffer $D$.\\
  \For{episode $0$ to max-episodes $T$,}{
      Initialize a random noise ${n_{act}}$ for exploring actions, and receive initial state ${{\bf{S}}_0}$, obtain each agents' observations from state $[o_1^i,o_2^i,...,o_t^i] \buildrel \Delta \over = {s_0}$.\\
      \For{t=1 to max-episode-length $T$,}{
          \For{each agent}{select actions from its observation and deterministic policy with exploration noise: $\bar a_j^i = clip({\bf{\tilde \pi }}_j^i(\tilde o_j^i) + \widetilde N(0,{e_{act}}),{{\bf{a}}_L},{{\bf{a}}_H})$}
          Execute action ${{\bf{a}}_j}$ and obtain new state ${\tilde s_j}$ and reward ${r_j}$.
          Store experience tuple $({s_j},{\tilde s_j},{{\bf{a}}_j},{r_j})$ into replay buffer $D$ and update ${s_{j + 1}} \leftarrow {\tilde s_j}$.
      }
      \If{episodes terminated and episodes > train-start-episode and episodes mod train-frequency,}{
          Randomly sample a mini-batch experience tuple $({{\bf{S}}_{sample}},{{\bf{\tilde S}}_{sample}},{{\bf{A}}_{sample}},{r_{sample}})$ from replay buffer $D$.
          Calculate target critic $Q$ value: ${y = r + \min {Q_{{{{\bf{\tilde \pi }}}^{1,2}}}}(({{\tilde O}^1},{{\bar A}^1}),...,({{\tilde O}^n},{{\bar A}^n})){|_{{{\bar A}^i} = {{{\bf{\tilde \pi }}}^i}({{\tilde O}^i}) + \widetilde N(0,{e_{critic}})}}}$.
          Update the value function by ${{\phi ^{1,2}} \leftarrow argmi{n_{{\phi ^{1,2}}}}\sum {{{({Q_{{{\bf{\pi }}^{1,2}}}}(({O^1},{A^1}),...,({O^n},{A^n})) - y)}^2}}}$.\\
          \If{episodes mod delay-frequency,}{
            Update all the policies ${{\nabla _{{\theta ^i}}}J({{\bf{\pi }}_i}) = \begin{array}{*{20}{c}}
{{{\mathop{\rm E}\nolimits} _{{\bf{O}},{\bf{A}} \sim D}}[{\nabla _{{\theta ^i}}}{{\bf{\pi }}^i}({A^i}|{O^i}){\nabla _{{A^i}}}{Q_{\bf{\pi }}}(({O^1},{A^1}),...,({O^n},{A^n}))]}\\
{_{{A^i} = {{\bf{\pi }}^i}({O^i})}}
\end{array}}$
            Soft-update target networks' parameters: $\begin{array}{l}
            {{\tilde \theta }^i} \leftarrow \tau {\theta ^i} + (1 - \tau ){{\tilde \theta }^i},\\
            {{\tilde \phi }^{1,2}} \leftarrow \tau {\phi ^{1,2}} + (1 - \tau ){{\tilde \phi }^{1,2}},
            \end{array}$
          }
      }
  }
\end{algorithm}

\subsection{Summary}
In this section, a self-attention-based multi-agent twin delayed deep deterministic policy gradient algorithm is proposed to deal with the uneven learning degree problem when the number of agents increases. The self-attention mechanism solves the problem of information fusion in an efficient way, and realizes the perspective-taking among agents. Its multi-input multi-output structure makes the method of value decomposition available. Three techniques organized around the self-attention mechanism are introduced that maximize its effectiveness: (1) The new input structure is the premise for the use of self-attention mechanism, and it shows the inherent relevance of the multi-agent and NLP problems. (2) The idea of value decomposition eliminates the ambiguity when value function is updated and makes the learning of multi-agents smoother. (3) The one-to-all policy update way improves the learning efficiency.

The aforementioned technique was combined with the MADDPG algorithm, named SA-MADDPG, and the double-attention-based algorithms (DSA-MADDPG and DSA-MATD3) were tested as well. This kind of double-attention structure cannot be decentralized execution. In many cases, projects that do not strictly require distributed execution are being worked on. In this case, such double-attention-based algorithms are recommended. In most experiments, they work better and the training is faster than that for single-attention-based algorithms. The pseudo-code based on the double-attention structure is given in the Appendix.
A comparison of the above-mentioned algorithm with other algorithms was also made.\\

\section{Experiment}
\label{sec:5}
\subsection{Environments}
Two typical scenarios in multi-agent particle environments (MPEs) were adopted to perform the experiments: cooperative-navigation and predator-prey scenarios, the details of which are provided below.

\begin{figure}[H]
    \centering
    \subfigure[Three agents]{\includegraphics[width=0.2\textwidth]{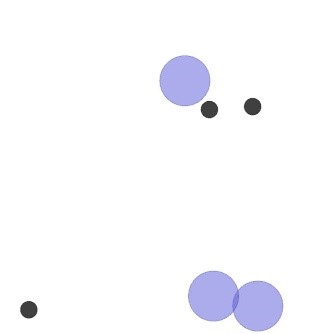}}
    \subfigure[Five agents]{\includegraphics[width=0.2\textwidth]{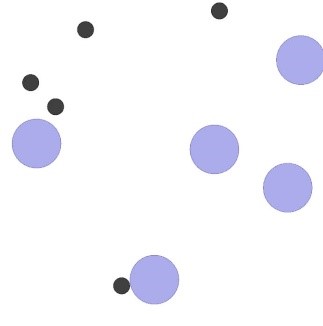}}
    \subfigure[Eight agents]{\includegraphics[width=0.2\textwidth]{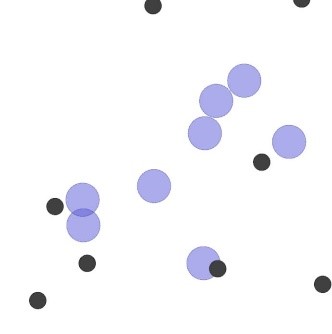}}
    \caption{Illustration of cooperative-navigation tasks.}
    \label{fig:Cooperative-Navigation}
\end{figure}

\textbf{Cooperative-Navigation Scenario}. In this scenario, the purple agents must cooperate through physical actions to reach a set of $L$ black landmarks. Agents observe the relative position of other agents and landmarks. In the experiment, the number of agents $N$ was set equal to the number of landmarks $L$. Agents are rewarded by the joint reward depends on their relative distance to the landmarks, and they learn to "cover" the landmarks that have not been "occupied" while avoiding other agents. To verify the performance of different algorithms with different numbers of agents, the number $N$ of agents in the scene was set to be three, five, and eight separately.

\begin{figure}[H]
    \centering
    \subfigure[Three agents]{\includegraphics[width=0.2\textwidth]{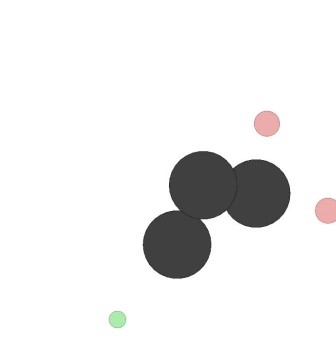}}
    \subfigure[Six agents]{\includegraphics[width=0.2\textwidth]{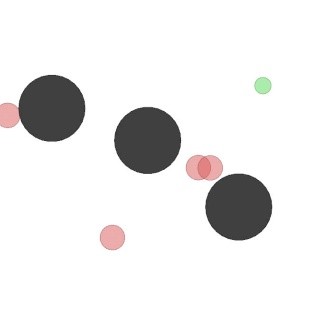}}
    \subfigure[Nine agents]{\includegraphics[width=0.2\textwidth]{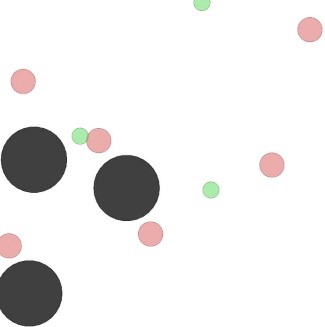}}
    \caption{Illustration of predator-prey tasks.}
    \label{fig:Predator-Prey}
\end{figure}

\textbf{Predator-Prey Scenario}. The red agents are the predators, the moving speed and acceleration of which are slower. The green agents are the prey, the moving speed and acceleration of which are slightly faster. In this scenario, $N$ slower cooperating agents must chase the faster adversaries. Once the predators collide with the preys, all of the predators will be rewarded, and the prey will be penalized at the same time.In addition, the prey cannot run out of the boundaries,or they will be rewarded negatively. The goal of the prey is to use their speed advantage and randomly generated obstacles in the environment to avoid the predators, and the predators should use their numerical advantage to chase the green agent. The number of predators was set as 2 times that of the prey. By loading pre-trained policies for prey, the predators' performance was compared under different algorithms. The number of agents $N$ was set to 3(2:1), 6(4:2), and 9(6:3) separately, and the number of obstacles $L$ is 3.

\subsection{Settings}
The proposed algorithms were evaluated with MADDPG and MATD3, which are MARL algorithms based on CTDE.
The SA-MATD3 is the algorithm proposed in this paper, while SA-MADDPG is an earlier experimental version that is based on MADDPG and can be regarded as an improved version of ATT-MADDPG. Self-attention was used instead of soft-attention, as were the various techniques mentioned in Section 3.4. DSA-MADDPG and DSA-MATD3 also use the self-attention mechanism in the actor network, and they cannot be decentralized execution. They are used as the reference in this paper. The learning curves are provided in the Appendix and the results summarized here.

\begin{table}[H]
	\caption{Comparisons with baseline algorithm}
	\centering
	\resizebox{\textwidth}{!}{
	\begin{tabular}{lllll}
		Name  & Remark  & Baseline algorithm   & Actor network, Critic network & Execution method\\
		\midrule
		MADDPG & 2017 & DDPG  & MLP,MLP   & Decentralized  \\
		MATD3 & 2020    & MADDPG & MLP,MLP & Decentralized     \\
		SA-MADDPG & Earlier version    & MADDPG    &MLP,Self-ATT   & Decentralized  \\
		DSA-MADDPG & Earlier version    & SA-MADDPG    &Self-ATT,Self-ATT   & Centralized  \\
		SA-MATD3 & This paper    & MATD3    &MLP,Self-ATT   & Decentralized  \\
		DSA-MATD3 & Modified    & SA-MATD3    &Self-ATT,Self-ATT   & Centralized  \\
		\bottomrule
	\end{tabular}}
	\label{tab:compare}
\end{table}
In the MPE, the actions are soft approximations to discrete messages, and Gumble-Softmax re-parameterization techniques\citep{gumble-softmax} are used to adapt to this environment.
All of the algorithms adopt the soft-target-update technique and share the same hyper-parameters, which is detailed in Appendix \ref{h_param}. They are tested on the same server and under the same environment configuration. A HP Z8 workstation with two Intel Xeon Gold 6242 CPUs and a Quadro GV 100 GPU were used. The operating system was Ubuntu 18.04 and the environment was CUDA 10.2. All of the algorithms were trained under Pytorch version 1.8.1.

\subsection{Experimental result in cooperative-navigation tasks}
\begin{figure}[H]
    \centering
    \subfigure[Reward]{\includegraphics[width=0.45\textwidth]{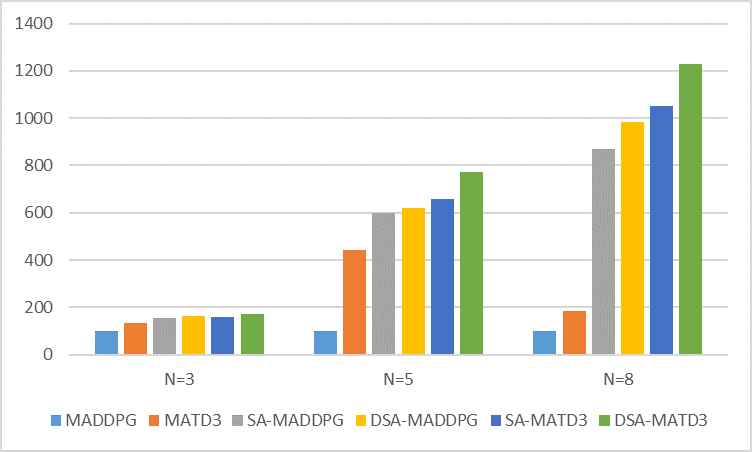}}
    \subfigure[Wall time]{\includegraphics[width=0.45\textwidth]{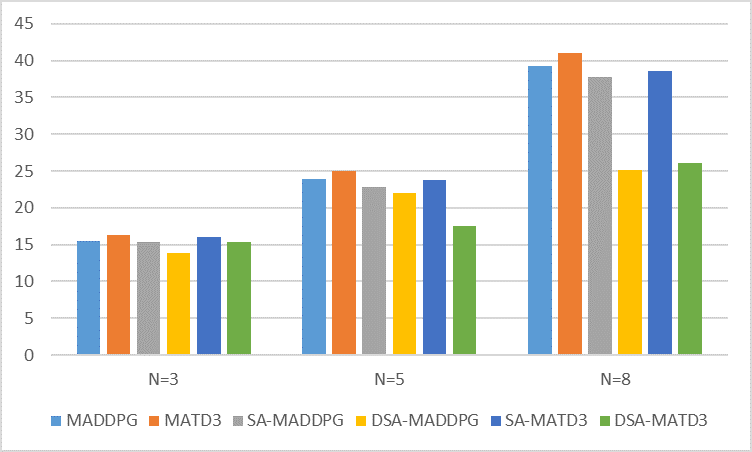}}
    \caption{Experimental results in cooperative-navigation scenario. Left, reward of different algorithms with different number of agents. Original reward is a negative value, so the reward value of MADDPG is normalized to 100 so that all of the rewards become positive values, which is more consistent with human cognition and easy to compare. Right, time spent by each algorithm in training 500,000 episodes.}
    \label{fig:sp}
\end{figure}

The cooperative-navigation scenario is examined first, and Figure \ref{fig:sp} shows the performance of various algorithms with different numbers of agents in that scenario.

When the number of agents is three, which is commonly used in most papers, there is no great difference in reward and wall time of various algorithms. When the number of agent is five, MADDPG can no longer learn the correct policy. When the number of agents increases to eight, MATD3 also fails. The performance of SA-MATD3 is second only to that of the centralized execution algorithm DSA-MATD3, and slightly better than that of the centralized execution algorithm DSA-MADDPG, which confirms that 
it is effective to use the "tricks" in TD3 to assist one-to-all updates. 

When one considers the training time, it can be seen that SA-MATD3 spends less than MADDPG and MATD3, In essence, a large part of time is spent in execution, since the centralized execution algorithm must only calculate the forward propagation once, while the decentralized execution algorithm must calculate the forward propagation $N$ times. A double self-attention-based algorithm spends much less time than other algorithms when there are a large number of agents. Comparison of wall time between a centralized execution algorithm and decentralized execution algorithm is meaningless because they have different policy networks, but the comparison of reward is still meaningful. Only the wall time of distributed execution algorithms is compared, and the training efficiency of the algorithm that takes less time is higher. SA-MATD3 takes less time than MATD3 and the gap widens as the number of agents increases.

\subsection{Experimental result in predator-prey tasks}
Because there are adversaries in this environment, they will be trained as the predator's policy changes, and due to this, the original environment is not conducive to the algorithm's comparison. Here, the pre-trained policies for prey are used\footnote{The prey are trained using TD3, and the predators using DSA-MATD3 as training partners; the pre-trained policies are changed every 1,000 episodes.} so as to achieve fairness in the algorithm comparison.

\begin{figure}[H]
    \centering
    \subfigure[Reward]{\includegraphics[width=0.45\textwidth]{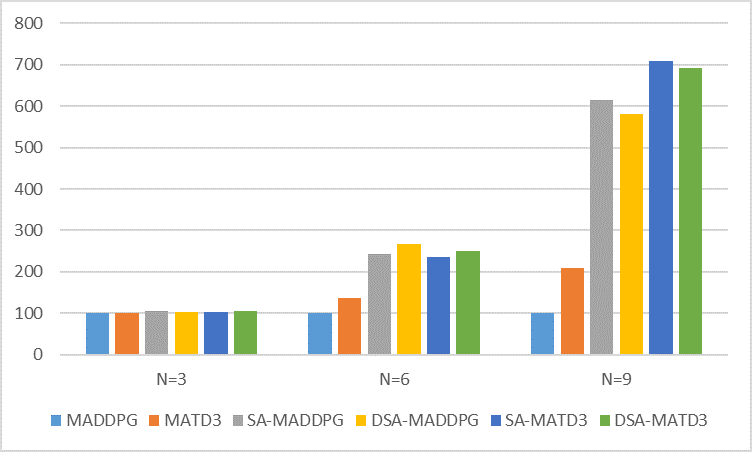}}
    \subfigure[Wall time]{\includegraphics[width=0.45\textwidth]{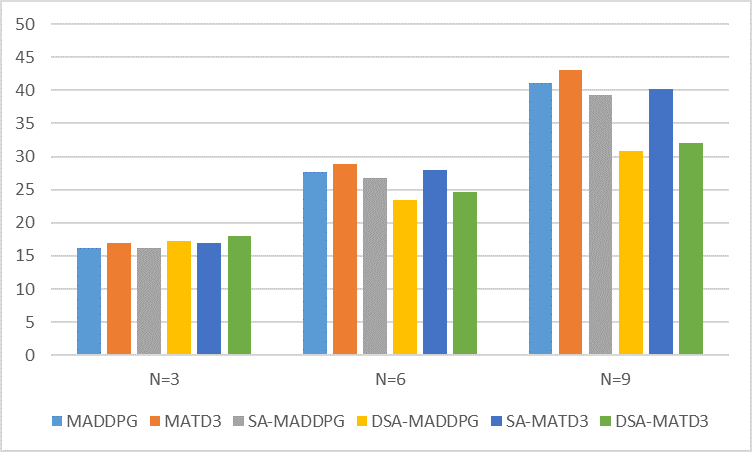}}
    \caption{Experimental results in predator-prey scenario. Left, reward of different algorithms with different number of agents; right, time spent by each algorithm for training 500,000 episodes. Here, reward value of MADDPG is normalized to 100 as well to make presentation clearer.}
    \label{fig:tag}
\end{figure}

In the predator-prey scenario, the reward of the predators is sparse, and it is more difficult for training compared to the cooperative-navigation scenario. As Figure \ref{fig:Predator-Prey} shows, the proposed algorithm is more efficient and effective when there are a large number of agents. In this scenario, the ratio of predator to prey is changed to 2:1; due to the reduction of collision probability, it is difficult for predators to learn the hunting policy based on the existing algorithms. Despite the time, SA-MATD3 can perform far better than the existing algorithms when the number of agents is six and nine, and even slightly better than the DSA-MATD3 when the number of agents is nine, which means that decentralized execution does not have inherent limitations compared with centralized execution. At the same time, it was found that when the number of prey is not one, the predators trained by the proposed algorithm can chase multiple prey, a phenomenon that is rarely shown by the existing algorithms.

\section{Conclusions and future work}
In this paper, the problem of uneven learning degree in multi-agent continuous control is raised, and self-attention-based multi-agent continuous control algorithms are proposed to solve it. The proposed algorithms could improve learning efficiency, realize perspective-taking among agents, and achieve superb performance when facing large-scale multi-agent tasks. 

Although the proposed algorithms have achieved exceptional results, they can only solve the cooperative problem at present. The transformer structure has proved to have great development potential. Encoding agents with different levels of performance will be considered in planned follow-up work in a way that is similar to positional encoding in an attempt to determine if the improved algorithms can handle large-scale multi-agent problems with cooperative as well as competitive tasks.

\section*{Acknowledgement}
This work was supported by the National Science Foundation of China under Grant No. 51975107, the National Natural Science Foundation of China under Grant No. 61371182, the Sichuan Science and Technology Major Project, No.2019ZDZX0020.
\bibliographystyle{elsarticle-num-names}
\bibliography{references}  






\appendix
\section{DSA-MATD3 algorithm}
\label{pscode}
\begin{algorithm}[H]
\caption{Double self-attention-based multi-agent twin delayed deep deterministic policy gradient.}
  Initialize actor network ${\bf{\pi }}$ with random parameters $\theta$, initialize two critic networks ${Q_{{{\bf{\pi }}^1}}},{Q_{{{\bf{\pi }}^2}}}$ with parameters ${\phi ^1},{\phi ^2}$, initialize target actor network $\tilde \theta  \leftarrow \theta $, initialize target critic networks ${\tilde \phi ^{1,2}} \leftarrow {\phi ^{1,2}}$,initial replay buffer $D$.\\
  \For{episode $0$ to max-episodes $T$,}{
      Initialize a random noise ${n_{act}}$ for exploring actions, and receive initial state ${{\bf{S}}_0}$, obtain each agents' observations from state $[o_1^i,o_2^i,...,o_t^i] \buildrel \Delta \over = {s_0}$.\\
      \For{t=1 to max-episode-length $T$,}{
          \For{each agent}{select actions from all agents' observations and deterministic policy with exploration noise: ${\bar a_j} = clip({{\bf{\tilde \pi }}_j}({\tilde o_j}) + \widetilde N(0,{e_{act}}),{{\bf{a}}_L},{{\bf{a}}_H})$}
          Execute action ${{\bf{a}}_j}$ and obtain new state ${\tilde s_j}$ and reward ${r_j}$.
          Store experience tuple $({s_j},{\tilde s_j},{{\bf{a}}_j},{r_j})$ into replay buffer $D$ and update ${s_{j + 1}} \leftarrow {\tilde s_j}$.
      }
      \If{episodes terminated and episodes > train-start-episode and episodes mod train-frequency,}{
          Randomly sample a mini-batch experience tuple $({{\bf{S}}_{sample}},{{\bf{\tilde S}}_{sample}},{{\bf{A}}_{sample}},{r_{sample}})$ from replay buffer $D$.
          Calculate target critic $Q$ value: ${y = r + \min {Q_{{{{\bf{\tilde \pi }}}^{1,2}}}}(({{\tilde O}^1},{{\bar A}^1}),...,({{\tilde O}^n},{{\bar A}^n})){|_{{{\bar A}^i} = {{{\bf{\tilde \pi }}}^i}({{\tilde O}^i}) + \widetilde N(0,{e_{critic}})}}}$.
          Update critic parameter by: ${{\phi ^{1,2}} \leftarrow argmi{n_{{\phi ^{1,2}}}}\sum {{{({Q_{{{\bf{\pi }}^{1,2}}}}(({O^1},{A^1}),...,({O^n},{A^n})) - y)}^2}}}$.\\
          \If{episodes mod delay-frequency,}{
            Update actor parameter ${{\bf{\pi }}^i}$ via policy gradient: $\begin{array}{l}
            {{\nabla _\theta }J({\bf{\pi }}) = \begin{array}{*{20}{c}}
{{E_{{\bf{O}},{\bf{A}} \sim D}}\left[ {\begin{array}{*{20}{c}}
{{\nabla _\theta }{\bf{\pi }}(({A^1}|{O^1}),...,({A^n}|{O^n})) \times }\\
{{\nabla _{{A^i}}}{Q_{\bf{\pi }}}(({O^1},{A^1}),...,({O^n},{A^n}))}
\end{array}} \right]}\\
{_{{A^1},{A^2},...,{A^n} = {\bf{\pi }}({O^1},{O^2},...,{O^n})}}
\end{array}}
            \end{array}$
            Soft-update target networks' parameters: $\begin{array}{l}
            \tilde \theta  \leftarrow \tau \theta  + (1 - \tau )\tilde \theta ,\\
            {{\tilde \phi }^{1,2}} \leftarrow \tau {\phi ^{1,2}} + (1 - \tau ){{\tilde \phi }^{1,2}},
            \end{array}$
          }
      }
  }
\end{algorithm}

\section{Hyper-parameters}
\label{h_param}
Different learning rates will have a great impact on the learning effect, so to ensure that each algorithm performs well, first, the performance of the two scenarios was tested at different learning rates, and the best-performing one taken as the learning rate for training. The smoothed rewards of MADDPG and DSA-MADDPG were compared over 10,000 episodes. The results are shown in Table \ref{tab:lr_1},\ref{tab:lr_2}.\\
\begin{table}[H]
    \caption{Results of different learning rates with five agents in cooperative-navigation environment}
    \centering
    \begin{tabular}{lll}
    \toprule
         lr & MADDPG & DSA-MADDPG  \\
    \midrule
         1e-2 & -2130 & -1747\\
         5e-3 & -1614 & -3560\\
         1e-3 & \bm{$-1585$} & -2207\\
         5e-4 & -1608 & -1757\\
         1e-4 & -1680 & \bm{$-1405$}\\
    \bottomrule
    \end{tabular}
    
    \label{tab:lr_1}
\end{table}

\begin{table}[H]
    \caption{Results of different learning rates with six agents in predator-prey environment}
    \centering
    \begin{tabular}{lll}
    \toprule
         lr&MADDPG&DSA-MADDPG  \\
    \midrule
         1e-2&0.5574&0.2736\\
         5e-3&0.4762&0.2523\\
         1e-3&\bm{$0.7825$}&0.2191\\
         5e-4&0.4612&0.2671\\
         1e-4&0.3754&\bm{$2.608$}\\
    \bottomrule
    \end{tabular}
    
    \label{tab:lr_2}
\end{table}
It can be seen from the experiment that the best learning rate of the MLP network is 1e-3 and that of the self-attention network is 1e-4 in both scenarios. Other hyper-parameters are listed in Table \ref{tab:hp}. This set of parameters was used in all of the experiments conducted in the present study.
\begin{table}[H]
    \caption{Hyper-parameters used in this paper}
    \centering
    \begin{tabular}{ll}
    \toprule
         Maximum number of steps per episodes&20\\
         Training start episodes&1e4\\
         Training frequency&5 episodes\\
         Batch size&512\\
         Standard deviation for action noise&0.002\\
         Standard deviation for critic noise&0.001\\
         Hidden dimensions&64\\
         Hidden layer&3\\
         Replay capacity&1e5\\
         Attention heads&4\\
         Optimizer &Adam\\
         Activation layer &Leaky Relu\\
         Negative slope &0.01\\
         Gamma&0.95\\
         Clipped grad&1\\
         Exchange rate for soft update&0.01\\
    \bottomrule
    \end{tabular}
    
    \label{tab:hp}
\end{table}

\section{Learning curves}
The experimental learning curves are shown in the following figures. Curves of episodes and wall time are plotted, respectively, to compare the performance and efficiency of different algorithms.
\begin{figure}[H]
    \centering
    \subfigure[Three agents]{\includegraphics[width=0.9\textwidth]{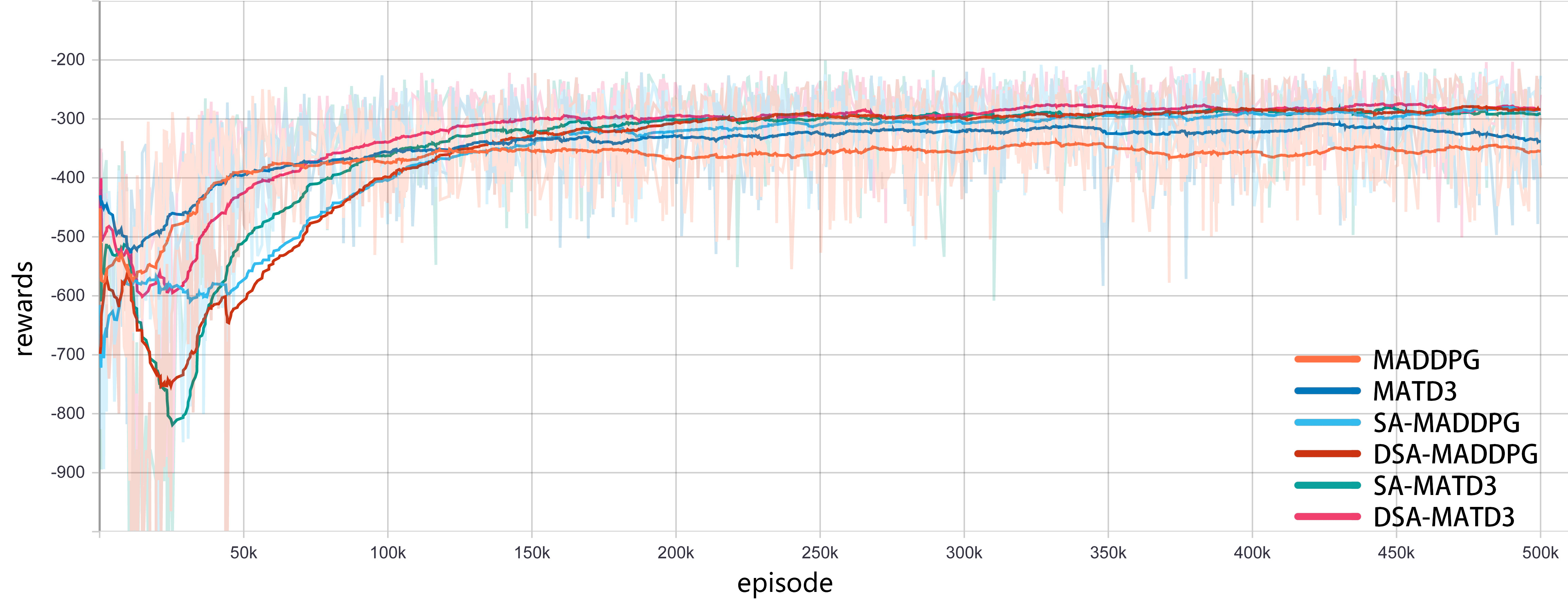}}
    \subfigure[Five agents]{\includegraphics[width=0.9\textwidth]{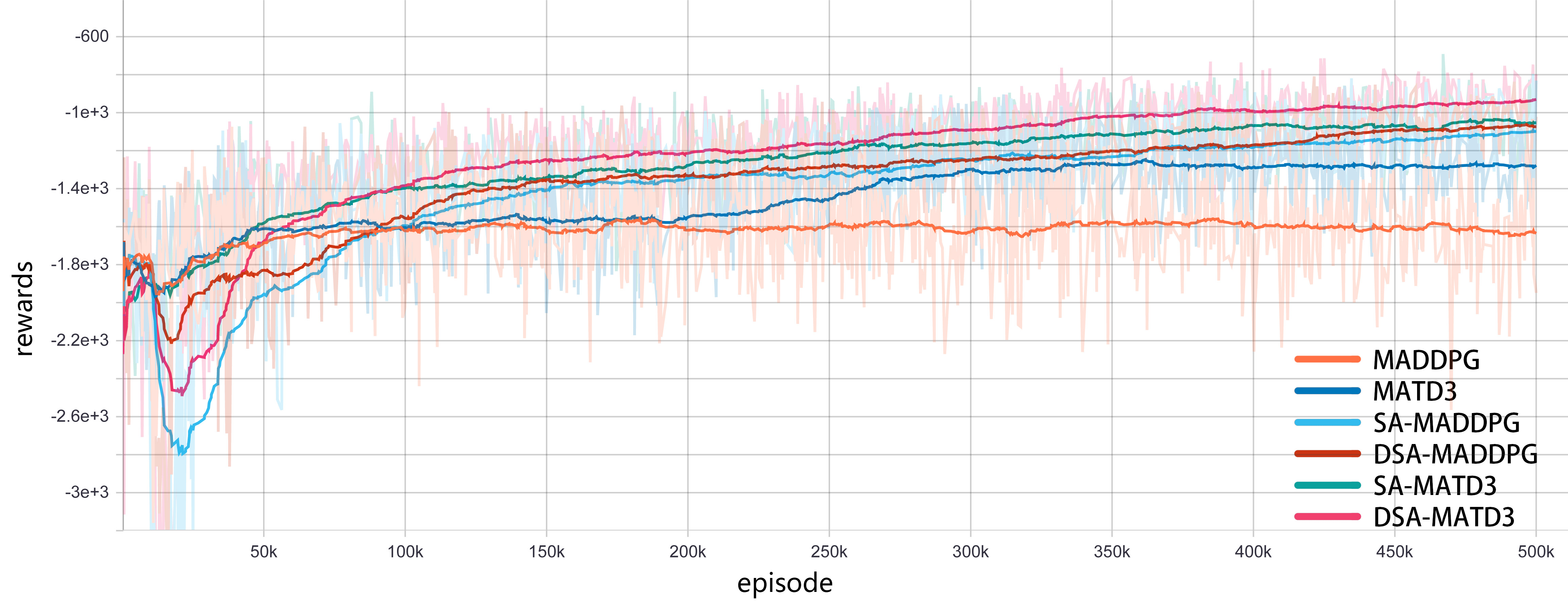}}
    \subfigure[Eight agents]{\includegraphics[width=0.9\textwidth]{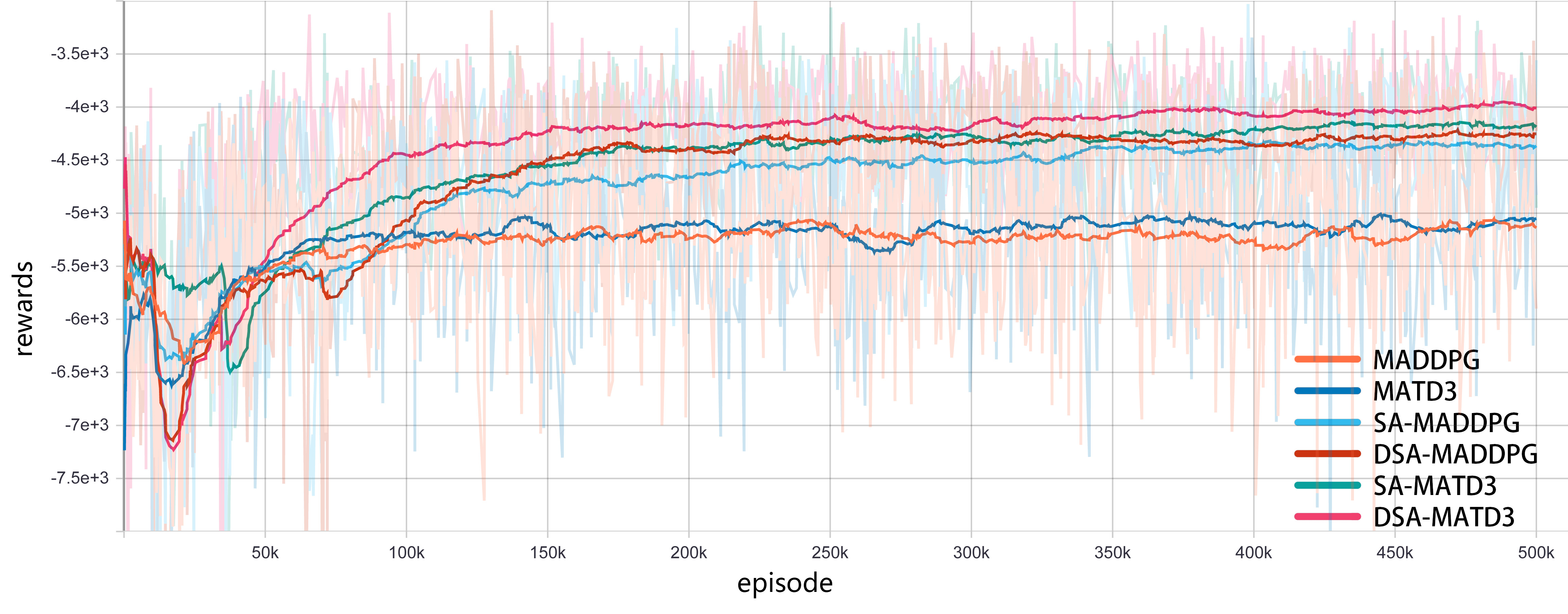}}
    \caption{Learning curves of episodes in cooperative-navigation scenario. It can be seen that with increasing number of agents the performance of MADDPG and MATD3 is poor, that of DSA-MATD3 is the best, and SA-MATD3 is slightly better than that of DSA-MADDPG.}
    \label{fig:tst1}
\end{figure}

\begin{figure}[H]
    \centering
    \subfigure[Three agents]{\includegraphics[width=0.9\textwidth]{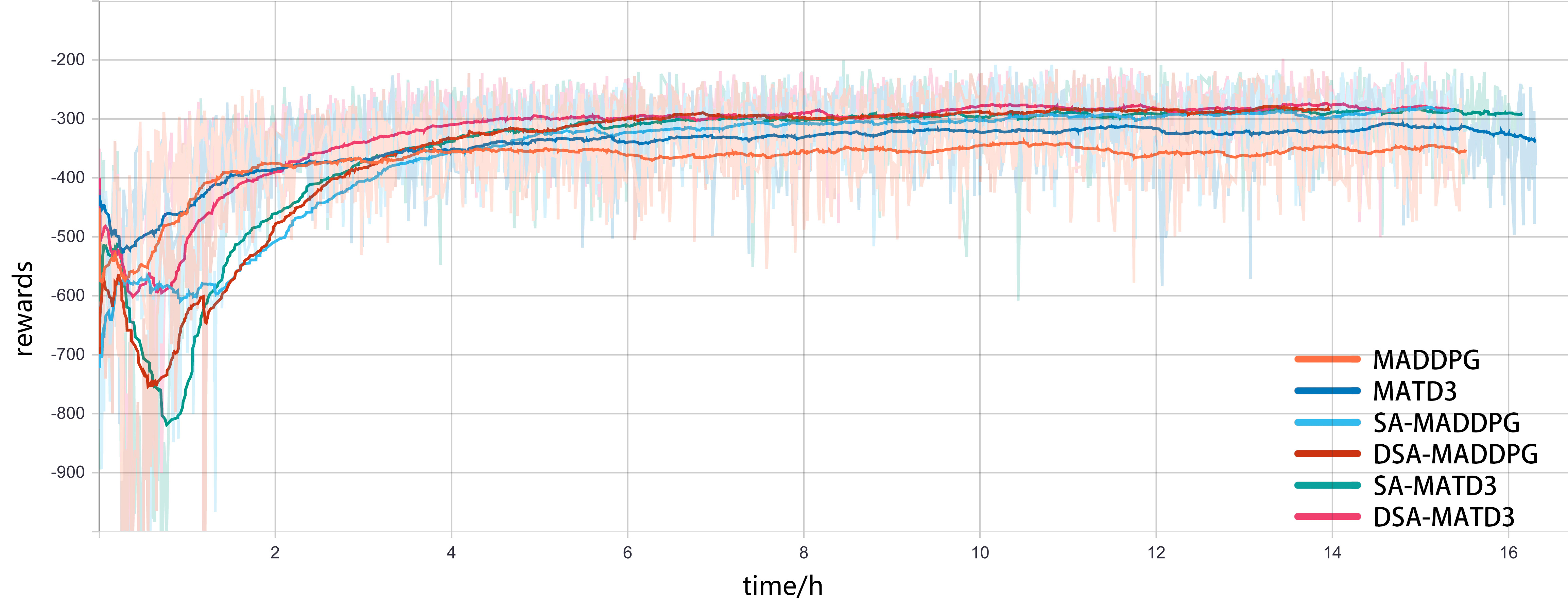}}
    \subfigure[Five agents]{\includegraphics[width=0.9\textwidth]{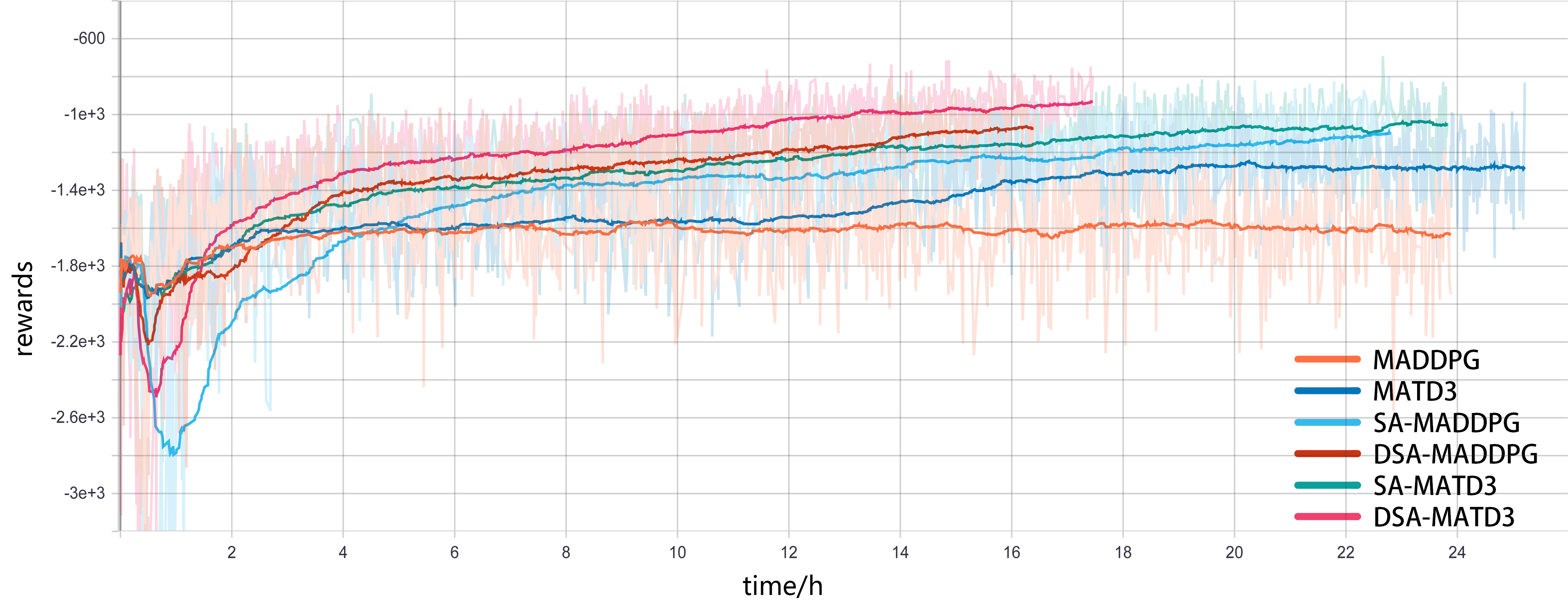}}
    \subfigure[Eight agents]{\includegraphics[width=0.9\textwidth]{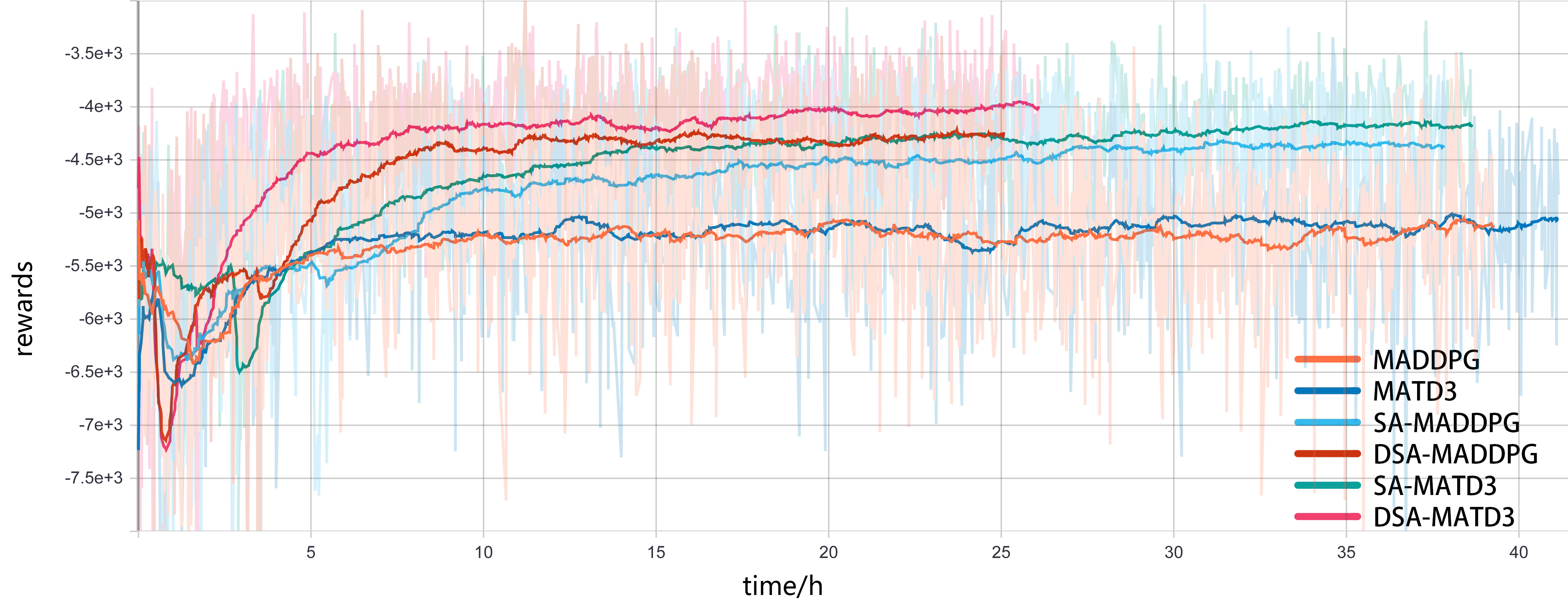}}
    \caption{Learning curves of wall time in cooperative-navigation scenario. Training of centralized algorithm is relatively fast, and this advantage is more obvious with increasing number of agents. Note that, although it does not seem that the relative time of proposed algorithm makes much of a difference compared with that of MADDPG or MATD3, the absolute time is increased with increasing number of agents.}
    \label{fig:tst2}
\end{figure}
 
\begin{figure}[H]
    \centering
    \subfigure[Three agents]{\includegraphics[width=0.9\textwidth]{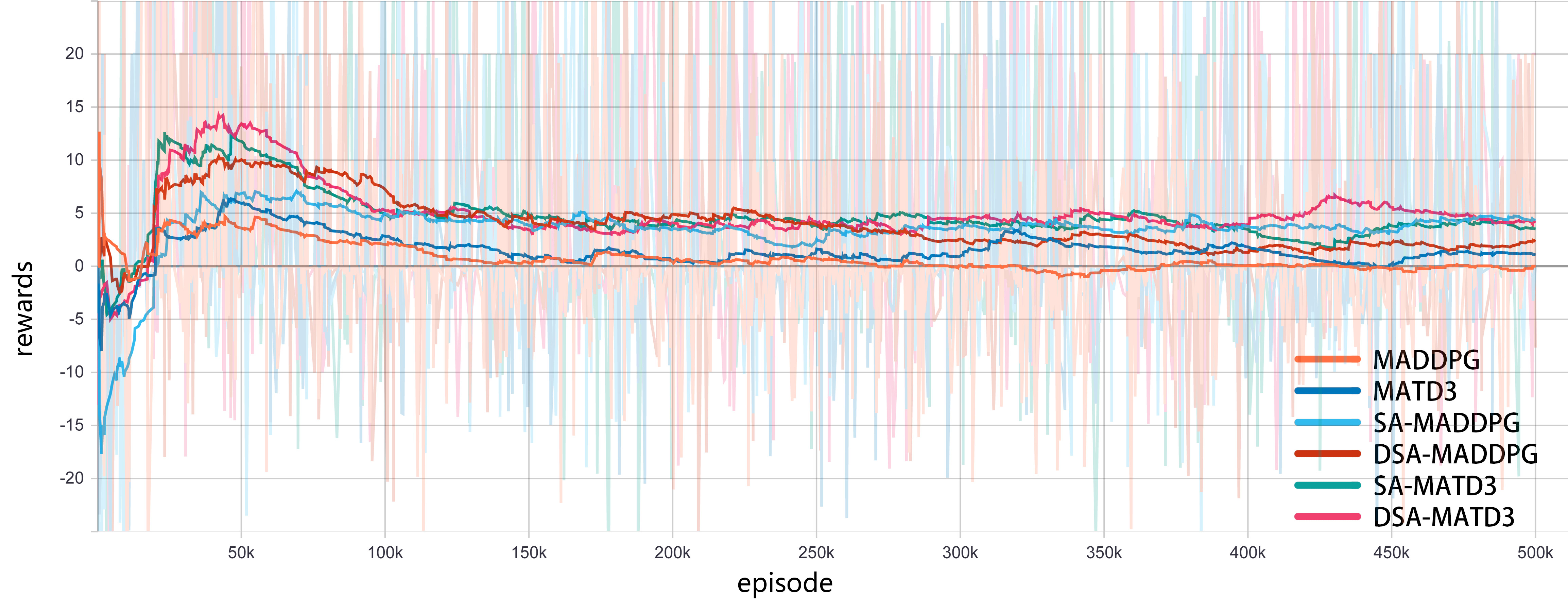}}
    \subfigure[Six agents]{\includegraphics[width=0.9\textwidth]{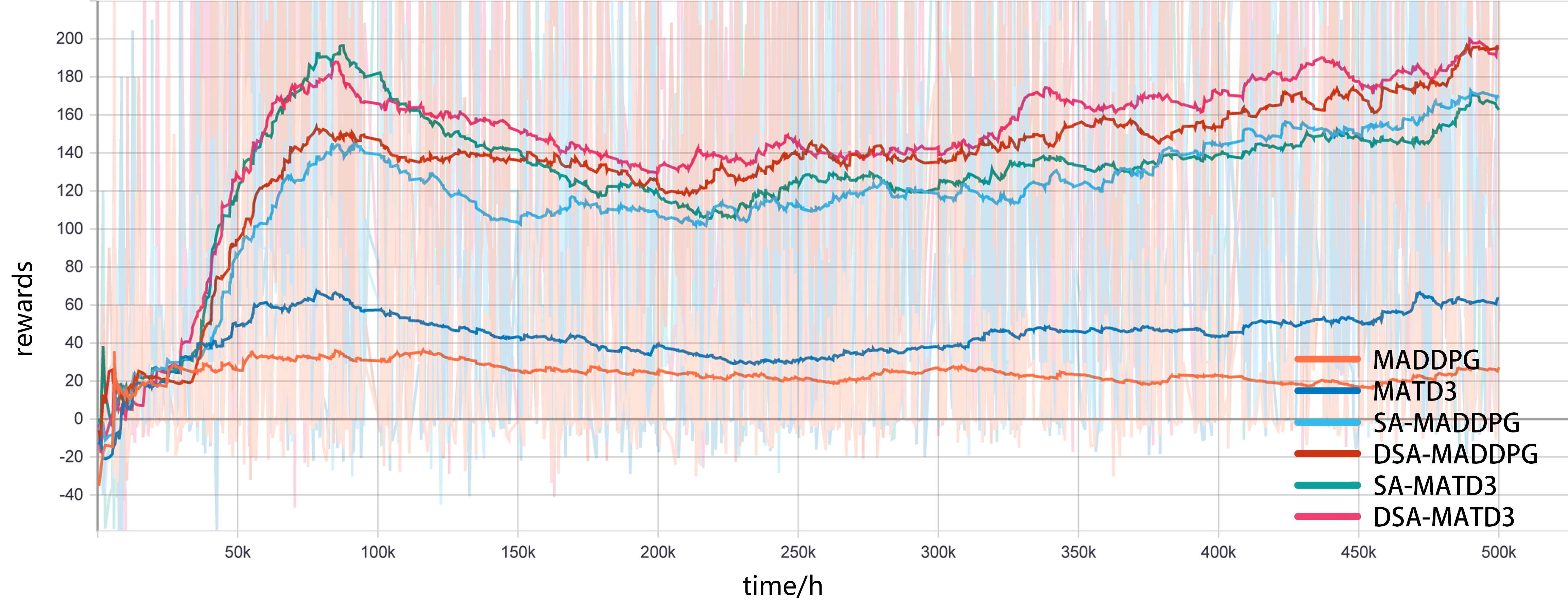}}
    \subfigure[Nine agents]{\includegraphics[width=0.9\textwidth]{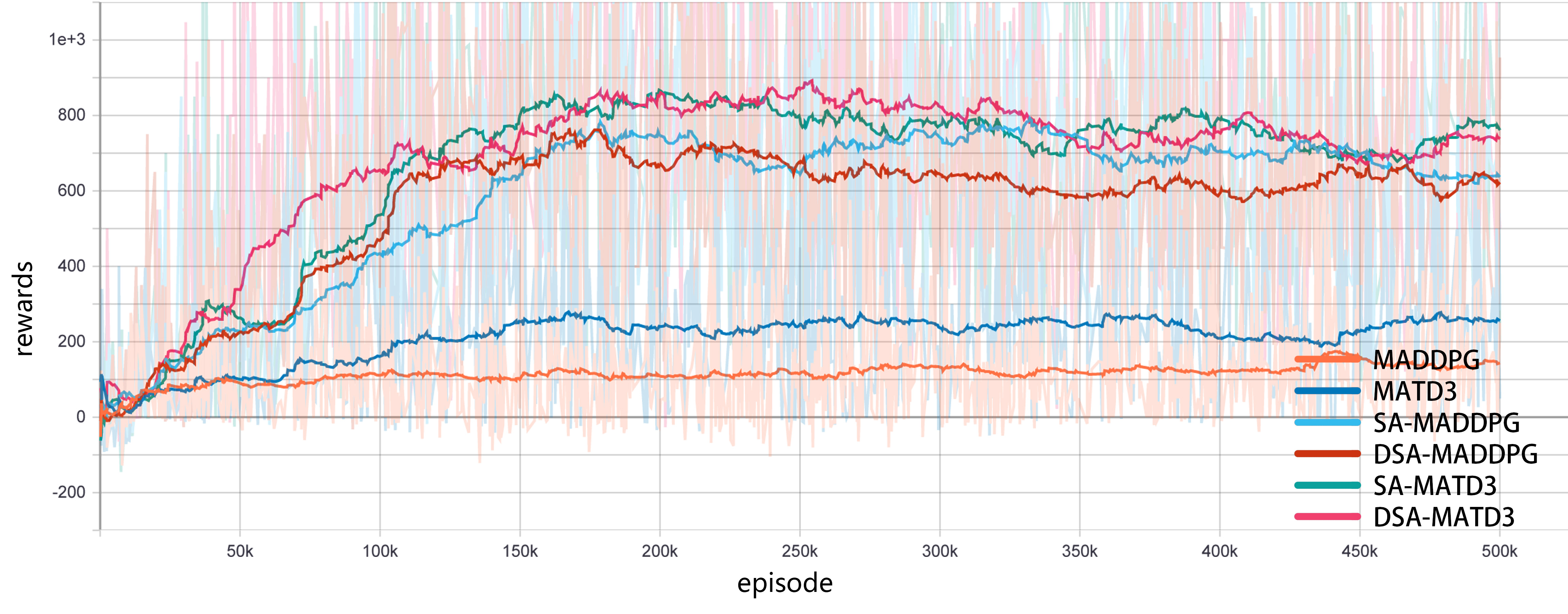}}
    \caption{Learning curves of episodes in predator-prey scenario. With increasing number of agents, gap between proposed and existing algorithm widens.}
    \label{fig:tst3}
\end{figure}

\begin{figure}[H]
    \centering
    \subfigure[Three agents]{\includegraphics[width=0.9\textwidth]{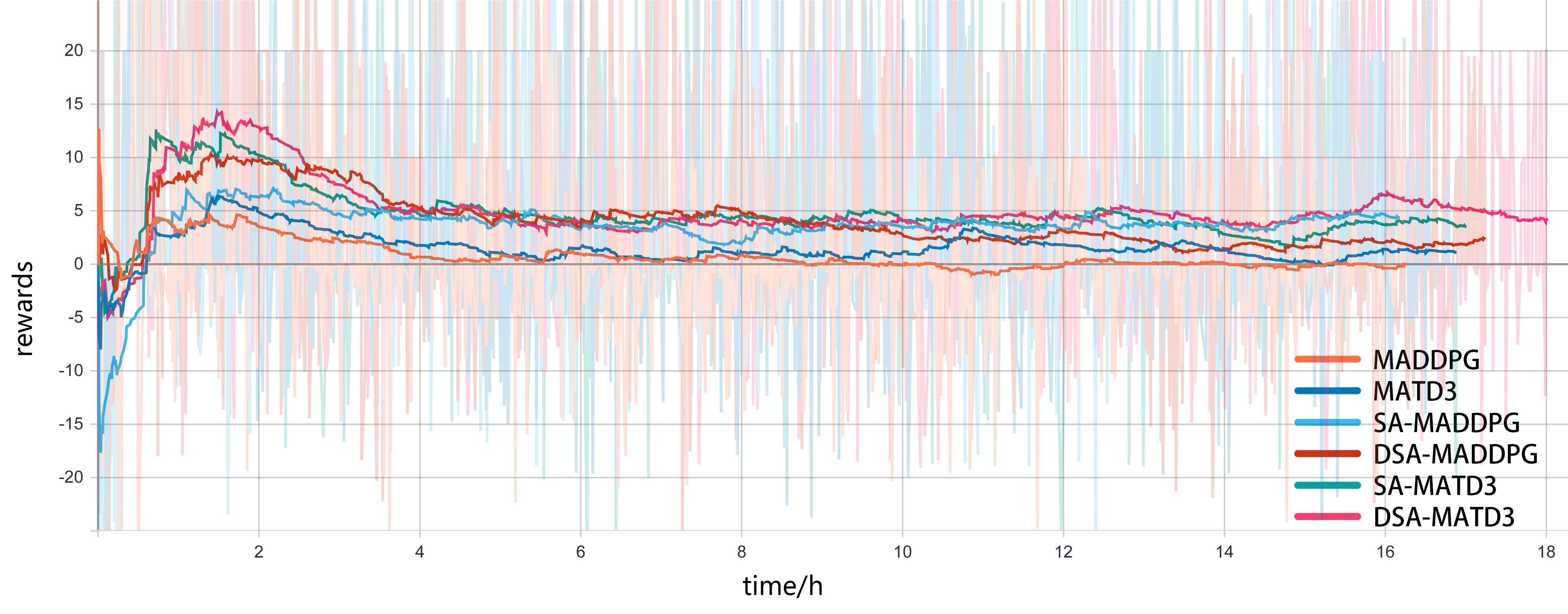}}
    \subfigure[Six agents]{\includegraphics[width=0.9\textwidth]{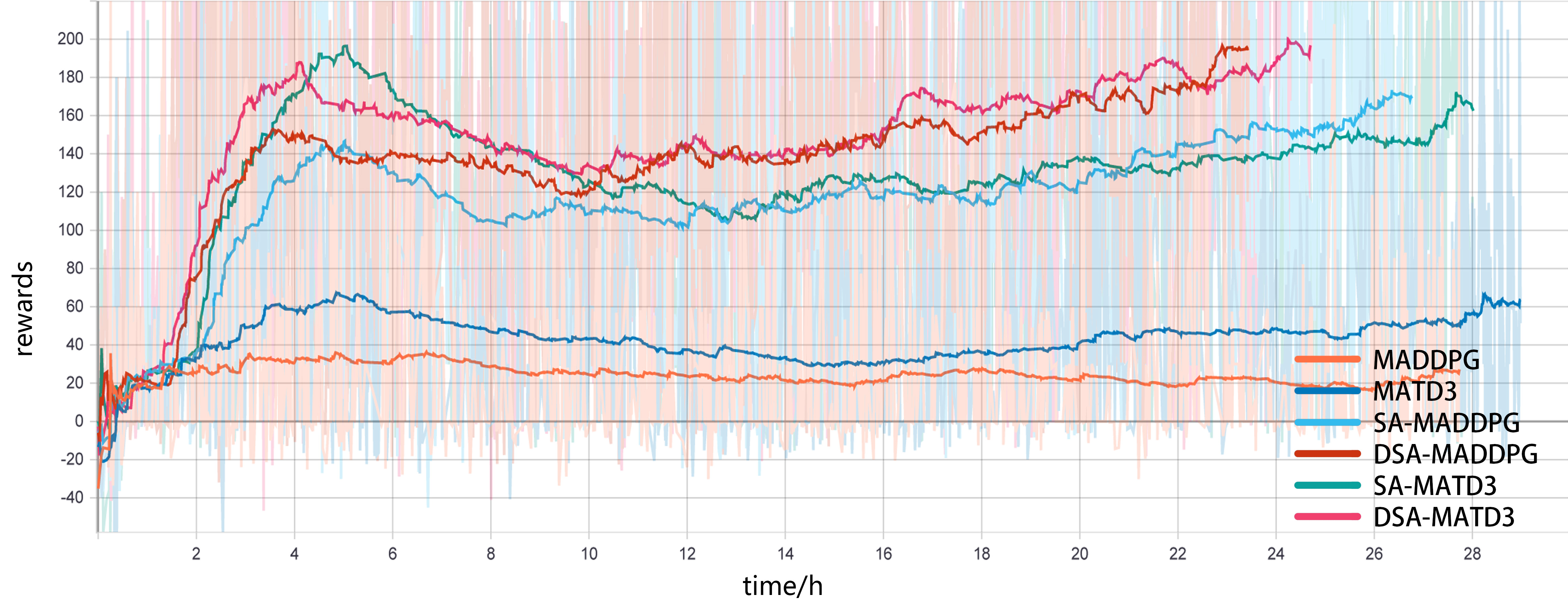}}
    \subfigure[Nine agents]{\includegraphics[width=0.9\textwidth]{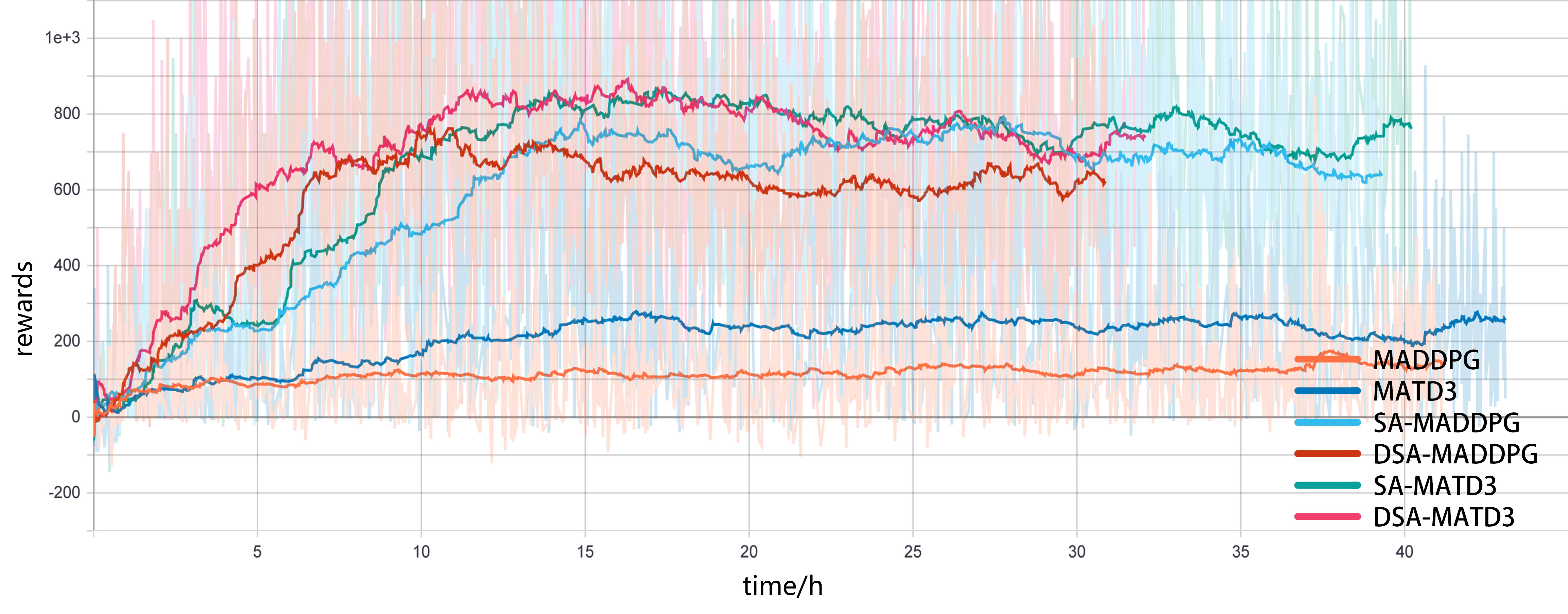}}
    \caption{Learning curves of wall time in predator-prey scenario. Both training efficiency and training effect of proposed algorithm are far superior to those of existing algorithms.}
    \label{fig:tst4}
\end{figure}






\end{document}